\documentclass{svproc}
\usepackage[table]{xcolor}
\usepackage{censor}
\usepackage{url, caption, graphicx, amsmath, soul, color}

\makeatletter
\renewcommand{\inst}[1]{\ifhmode\unskip\fi$^{#1}$}
\makeatother

\definecolor{myred}{rgb}{0.9, 0.0, 0.0}
\definecolor{darkred}{rgb}{0.7, 0.0, 0.2}
\definecolor{redfill}{rgb}{1.0, 0.41, 0.38}
\definecolor{mygreen}{rgb}{0.0, 0.5, 0.0}
\definecolor{greenfill}{rgb}{0.66, 0.89, 0.63}
\definecolor{myblue}{rgb}{0.0, 0.2, 0.7}
\definecolor{darkblue}{rgb}{0.2, 0.2, 0.7}
\definecolor{mygrey}{rgb}{0.5, 0.5, 0.5}

\usepackage{afterpage, booktabs, multirow}

\begin{document}
\mainmatter              
\title{Lightweight Hopfield Neural Networks for Bioacoustic Detection and Call Monitoring of Captive Primates}
\titlerunning{Lightweight Associative Memory Models for Bioacoustic Detection}  
%
\author{Wendy Lomas\inst{1} \and Andrew Gascoyne\inst{2} \and Colin Dubreuil\inst{3} \and Stefano Vaglio\inst{4} \and Liam Naughton\inst{5}}
\authorrunning{Lomas et al.} 
\institute{\inst{1,2,5}School of Engineering, Computing and Mathematical Sciences, \inst{3,4}School of Pharmacy and Life Sciences, University of Wolverhampton, Wolverhampton, UK,\\
\email{\{w.k.lomas, a.d.gascoyne, c.dubreuil, s.vaglio, l.naughton\}@wlv.ac.uk}
}


\maketitle  

\begin{abstract}
Passive acoustic monitoring is a sustainable method of monitoring wildlife and environments that leads to the generation of large datasets and, currently, a processing backlog. Academic research into automating this process is focused on the application of resource intensive convolutional neural networks which require large pre-labelled datasets for training and lack flexibility in application. We present a viable alternative relevant in both wild and captive settings; a transparent, lightweight and fast-to-train associative memory AI model with Hopfield neural network (HNN) architecture.  Adapted from a model developed to detect bat echolocation calls, this model monitors captive endangered black-and-white ruffed lemur (\textit{Varecia variegata}) vocalisations.  Lemur social calls of interest when monitoring welfare are stored in the HNN in order to detect other call instances across the larger acoustic dataset.  We make significant model improvements by storing an additional signal caused by movement and achieve an overall accuracy of $0.94$.  The model can perform $340$ classifications per second, processing over $5.5$ hours of audio data per minute, on a standard laptop running other applications. It has broad applicability and trains in milliseconds. Our lightweight solution reduces data-to-insight turnaround times and can accelerate decision making in both captive and wild settings.

\keywords{Artificial intelligence, Hopfield network, bioacoustics, passive acoustic monitoring, associative memory, neural networks, signal processing, black-and-white ruffed lemurs, Varecia variegata}
\end{abstract}

\section{Introduction}\label{sec:intro}

Within the fields of conservation and ecology, sustainable methods of monitoring and surveying animal populations are vital to help arrest biodiversity loss and alleviate climate change.  Bioacoustics offers an innovative approach to biodiversity measurement and in this context refers to the study of the sounds produced by non-human animals and how these sounds are transmitted and received.  One method of bioacoustic event collection is passive acoustic monitoring (PAM), where autonomous acoustic recording units are placed in the field and the collected data are analysed.  These acoustic workflows are increasingly being used for ecological surveying and conservation \cite{bak}; however, there is currently a data processing backlog \cite{ker:etal}.  Manual bioacoustic analysis is labour intensive and automated approaches, such as commercial classifiers, have limited precision and applicability \cite{mar:fab:aub,tab:etal}. Therefore, there is a pressing need for new reliable and versatile automatic analysis tools that can be quickly implemented \cite{ker:etal}. The focus amongst researchers is on the application of convolutional neural networks (CNNs) to automate bioacoustic event detection using spectrograms \cite{ker:etal,ras:sto:bri,cau:fav:mar:rey,bat:etal}.  These approaches can lack transparency for field workers and are computationally expensive, as they require extensive pre-labelled datasets for training and the conversion of raw sound files to images in the form of a spectrogram with specific dimensions required by the particular CNN architecture.  Spectrograms are produced from the sound file by discretising the signal from a window in the time domain to the frequency domain via a fast Fourier transform (FFT) and then reconstructing it to show frequency vs. time with amplitude represented by intensity.  Clark and Dunn \cite{cla:dun} review the use of acoustic monitoring in captive environments and suggest that less commercial settings, such as zoos, could greatly benefit if able to secure the funding and expertise required for implementation and analysis. PAM has the potential to help measure the impact of welfare interventions, initiatives to enhance reproduction, enclosure changes, and noise across the zoo site on resident animals; this list is by no means exhaustive but gives a good sense of the range of insights that could be gained from acoustic monitoring. However, zoos as critical centres for conservation research have not yet been able to maximise the advances evidenced in the bioacoustic literature base \cite{cla:dun}, with both finance and applicability of the models being developed for the field playing a factor in this lack of uptake in captive settings. 

In this paper, we develop a model to track captive black-and-white ruffed lemur (\textit{Varecia variegata}) calls. The Critically Endangered \cite{iuc} black-and-white ruffed lemur, was the subject of an undergraduate wildlife conservation project in collaboration with a zoological garden (see section \ref{sec:ack}) in 2023, where an audio dataset was collected for later analysis.  Two female lemurs and one male were housed in an enclosure with access to outdoor space, and after some experimentation an AudioMoth (AM) \cite{hil:etal} was placed in the indoor enclosure and over $30$\,hours of audio data collected in a period of one month.  Two calls were identified to be tracked across this dataset, ``grumbles'' and ``alarms'', and a portion of the dataset labelled in order to evaluate the model developed based on \cite{gas:lom}.  In \cite{gas:lom} the first-of-its-kind bioacoustic event detection model via a Hopfield network architecture was proposed and developed to identify bat echolocation calls.  Hopfield neural networks (HNNs) \cite{hop} are fully connected recurrent neural networks that simulate biological associative memory by storing patterns and retrieving these patterns from partial or noisy inputs, leading to their use as content-addressable memory systems.  An HNN is an example of an undirected graph, as the weights between neurons are symmetrical and thus simulate communication between biological neurons and within neural circuits.  HNNs are dynamical models based on zero temperature Ising models \cite{isi} where the energy of the system is defined by a function which converges towards local minima on every network update.  As described in \cite{hop:tan} and due to this evolution of the network to energy minima, if an optimisation problem can be written in the form of a Hopfield energy function, then the minima of the HNN represent solutions to that problem.  Hopfield networks have been further improved for optimisation problems by the inclusion of transient chaotic neuronal dynamics in order to efficiently span the energy landscape and hence discover the global minimum of optimisation problems \cite{rod:etal,che:aih,xu:zha}.  However, the deep learning models most commonly used in the literature for bioacoustic classification are CNNs, feed-forward deep neural networks with convolutional layers ideal for image classification tasks.  Unfortunately, like most deep learning models, CNNs require large amounts of training data and are resource intensive.  We present a lightweight associative memory alternative with HNN architecture and with broad applicability in captive settings, able to classify raw sound files of any length.  We demonstrate here that these models, proven to be accurate classifying bat echolocation pulses in relatively uncluttered high frequency soundscapes \cite{gas:lom}, can be adapted for the harder task of detecting primate vocalisations in the cluttered lower frequency range audible to humans.  Furthermore, we demonstrate that the models can be built rapidly to track specific vocalisations within a few hours of receiving a dataset and the calls to be detected.

We organise the paper as follows. The methodology is described in section \ref{sec:method}, including the collection and labelling of the audio dataset in section \ref{sec:data} and a full description of the theory and architecture of the associative memory models used in section \ref{sec:models}.  In sections \ref{sec:results} and \ref{sec:discussion} we examine the classifications generated by the model after testing with the labelled dataset, and discuss how to make further improvements before summarising and drawing our final conclusions about the implications for conservation and zoo husbandry in section \ref{sec:conc}.

\section{Methodology}\label{sec:method}

In this section, we first discuss the dataset of audio files collected in an enclosure housing black-and-white ruffed lemurs at a zoological garden in November/December 2023.  We then outline the theory behind the associative memory models used and how they were then developed and implemented for this dataset.

\subsection{Dataset}\label{sec:data}

\begin{figure}[ht]
\centering
\includegraphics[width=0.95\textwidth]{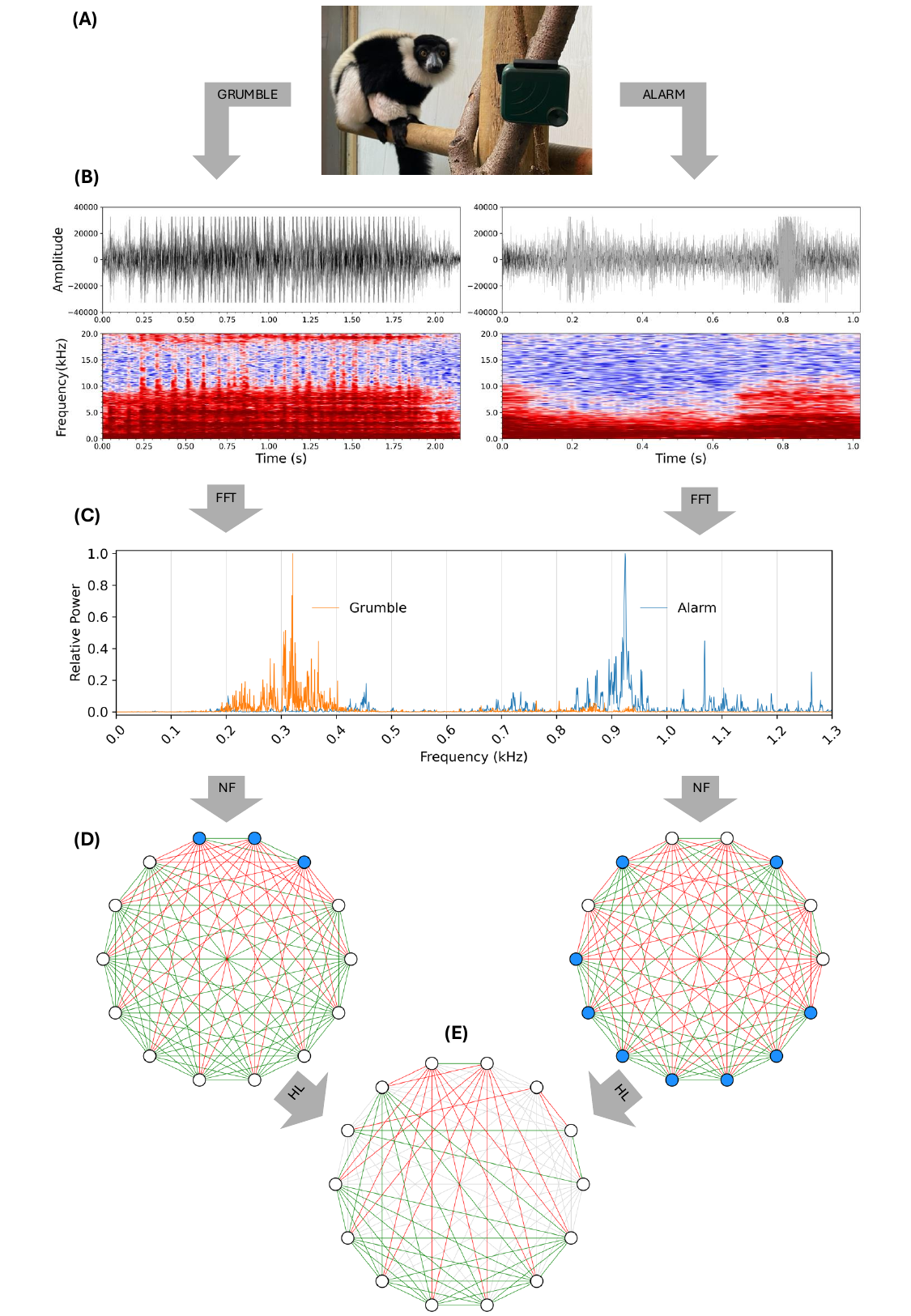}
\caption{Model 1 training: sound files are collected (A); two specific calls to be monitored are identified (B); the fast Fourier transform (FFT) is applied to these signals and peak frequencies extracted from the frequency range of interest (C); neurons fire (NF) based on the FFT peaks and the network is activated (\textcolor{myblue}{\bf blue}) for each signal (D); these network activations are then combined via Hebbian learning (HL) (equation \ref{eqn:Hebb}) and the trained model is ready for activation/classification (E).  \textcolor{mygreen}{\bf Green} and \textcolor{myred}{\bf red} lines represent positive and negative neuron associations (weights) respectively and \textcolor{mygrey}{\bf grey} lines represent disassociated neurons (zero weights).
}\label{fig:flow1_activ}
\end{figure}

\begin{figure}[ht]
\centering
\includegraphics[width=0.95\textwidth]{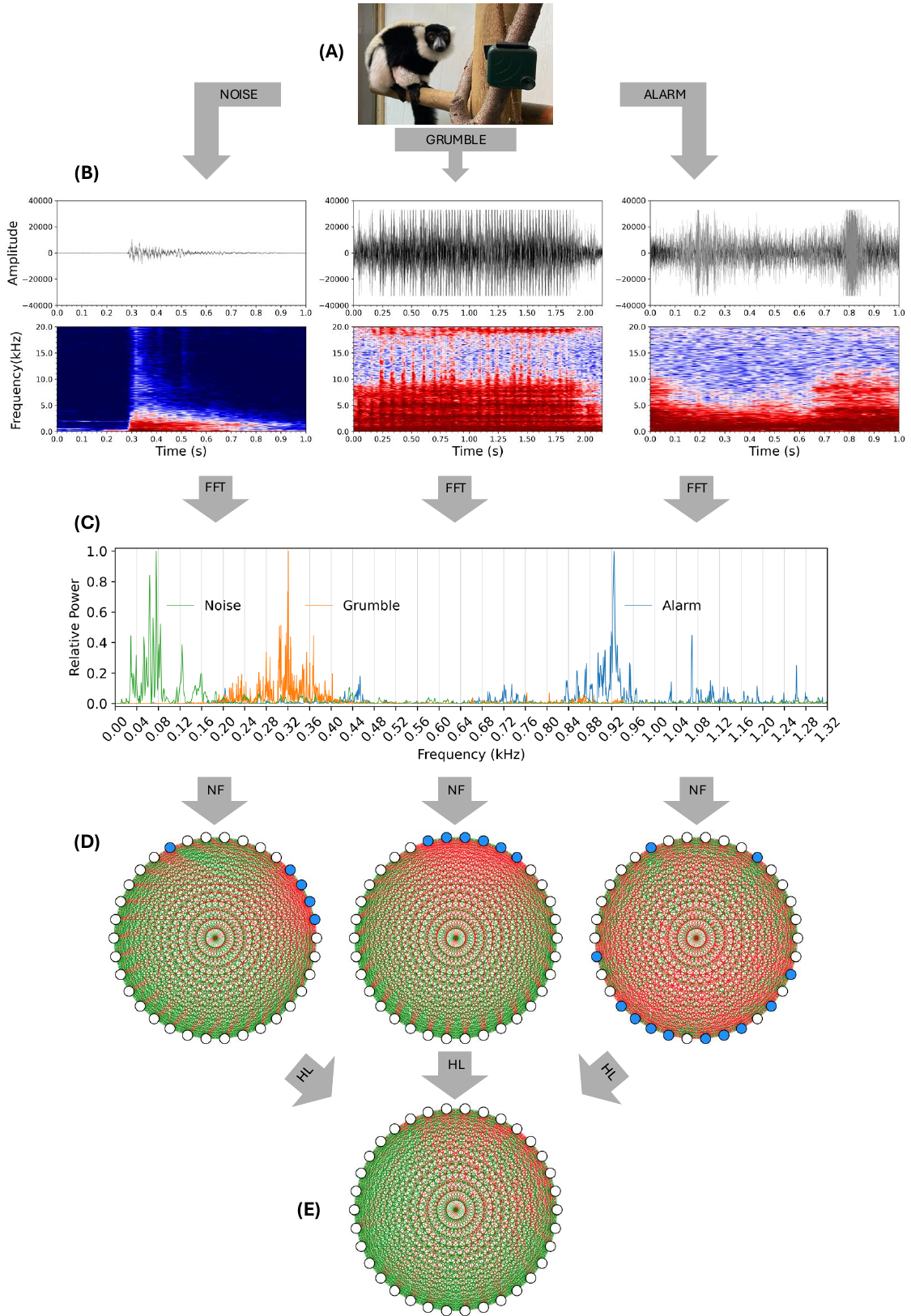}
\caption{Model 2 training: signals are collected (A); two specific calls to be monitored are identified (B) and one ``noise'' to be filtered; the FFT is applied to each signal and peak frequencies extracted (C); neurons fire (NF) based on the FFT peaks and the network is activated (\textcolor{myblue}{\bf blue}) for each signal to be stored (D); network activations are combined by Hebbian learning (HL) (equation \ref{eqn:Hebb}); the trained model is ready for activation/classification (E), where \textcolor{mygreen}{\bf green} and \textcolor{myred}{\bf red} lines represent positive and negative neuron associations (weights) respectively and \textcolor{mygrey}{\bf grey} lines represent disassociated neurons (zero weights).
}\label{fig:flow3_activ}
\end{figure}

The Critically Endangered \cite{iuc} black-and-white ruffed lemur, was the subject of an undergraduate wildlife conservation project in collaboration with a zoological garden in 2023, where an audio dataset was collected for later analysis.  The project was part of a wider scent enrichment programme inspired by the work of \cite{elw:vag,fon:etal}.  Two female black-and-white ruffed lemurs and one male were housed in an enclosure with access to outdoor space.  After experimenting with the location of the AM recorders \cite{hil:etal}, the student and keepers settled on an indoor installation attached to a branch within the enclosure; see figure \ref{fig:flow1_activ} (A). More than $30$\,hours of audio data were collected over a period of one month on seven non-consecutive days. This comprised $73$\,.WAV formatted files with a maximum length of $1795$\,seconds and a $5$\,second sleep period per file to allow recording to the SD card. The AudioMoths were configured and turned on before setup, and turned off after removal. Setup and removal occurred on each day of recording; subsequently, the setup and removal periods were removed from the dataset as these did not contain representative or relevant recordings; see sections \ref{sec:results} and \ref{sec:discussion} for a full discussion of other dataset limitations and recommendations.  After manual examination of the sound files and discussions with the conservation team and primate lead at the zoo, two call types of interest were selected, ``alarms'' and ``grumbles''.  The signals and spectrograms are shown in figure \ref{fig:flow1_activ} (B).  Concern about the recorder location and noise from lemur movement meant that a third ``noise'' sound file was also identified as shown in figure \ref{fig:flow3_activ} (B). This was representative of the sound collected as the lemurs moved around and on the branch to which the recorder was connected.

In consultation with the zoo team we named the captive lemur calls of interest ``alarms'' and ``grumbles'' for the purposes of this study and in line with the Duke Lemur Center \cite{goo}, a non-invasive research centre housing the most diverse captive population of lemurs in the world. The alarm calls in our study are similar to the calls characterised as roar-shrieks in a population of wild Madagascan black-and-white ruffed lemurs where $11$ call types are identified and where it is noted that calls are often used in combination as syllables and in chorus \cite{bat:raz:ran:bad}. To limit cross-over with other vocalisation types within the repertoire of the three captive lemurs studied in this paper, we describe bouts of alarms and grumbles as follows:  
\begin{itemize}
\item Alarm bouts are sequences of alarm calls of $3$\,seconds or more in duration, often triggered in response to hearing other lemurs in neighbouring enclosures or to helicopters and, as such, were characterised as territorial and/or warning vocalisations by the team at the enclosure.  
\item Lower frequency grumble bouts are sequences of grumble calls of $2$\,seconds or more in duration and are individual vocalisations considered by the enclosure team to be anticipatory social calls made whilst the lemur is looking down and then followed by the lemur looking up and around.
\end{itemize}
Representative sounds of these two calls were found in one audio file collected on the first day and short samples of these were saved to be used for model training as described in section \ref{sec:models} and visualised in figure \ref{fig:flow1_activ} (B).  We should remember that here neither a perfect sound nor many examples are required; the aim is to achieve a working model quickly and with applicability across the wider dataset.

Data collected on the first $4$\,days of the study were labelled manually in order to evaluate the performance of the models developed in section \ref{sec:models}.  To complete this manual labelling process as efficiently as possible, the files were split into $1064$\,one-minute signals and viewed on a bespoke spectrogram visualiser and audio player.  A total of $203$\,grumble bouts were manually found and labelled, along with $32$\,alarm bouts.  This was consistent with the expectations of the primate lead at the zoo; grumbles occur frequently throughout the day whereas alarms may only occur a few times per day.  Other signals are labelled as ``non-call'' bouts. These last between $1$ and $60$\,seconds and are signals not considered to be either ``alarms'' or ``grumbles''; there are $1263$ non-call bouts within the dataset.  This manual labelling process took around $10$\,hours to complete, with a similar amount of time being spent to manually check the model classifications; see section \ref{sec:results}.  The models used to automate call detections and make these classification are described in the following section.

\subsection{Associative Memory Models for Bioacoustic Detection of Lemur Social Calls}\label{sec:models}

\afterpage{\clearpage}

\afterpage{\clearpage}

The model developed here was adapted from \cite{gas:lom} and simulates and augments the human hearing process to accelerate bioacoustic analysis and enhance existing conservation methods. The model consists of a Hopfield neural network \cite{hop}, which is an example of associative or content-addressable memory, to store signals as patterns and then detect and classify similar signals.  Biologically inspired Hopfield neural networks (HNNs) are examples of fully connected recurrent neural networks with no separate input or output neurons. A discrete HNN is applied where: neuron firing is constrained to a binary output; weights between neurons are symmetric, $w_{ij}=w_{ji}$; and neurons are not self-connected, $w_{ii}=0$.  When sounds are stored to the network they are known as retrieval states, and network weights are set according to Hebbian learning \cite{heb}:
\begin{equation}\label{eqn:Hebb}
\mathbf{W}=\frac{1}{N}\sum_{k=1}^p\mathbf{X}^k\cdot\mathbf{X}^k
\end{equation}
where each $\mathbf{X}^k$ is a retrieval state representing the stored signals, $p$ is the number of stored signals and $N$ is the number of neurons in the network. Hebbian learning works on the principle that neurons that fire together, wire together and therefore future activity of a neuron will affect its associated neurons.  Once the network is established we can present a pattern to the network whose neurons, governed by dynamical equations \cite{gas:lom}, evolve and converge to a local minima of the system's energy function defined by \cite{hop:tan}:
\begin{equation}\label{eqn:Energy}
E=-\frac{1}{2}\sum_i\sum_j w_{ij}x_ix_j-\sum_i I_ix_i
\end{equation}
where $E$ is the system's energy, $I_i$ is the input bias of neuron $i$, $w_{ij}$ is the weight between neuron $i$ and neuron $j$, and $x_i$ refers to the binary state of the $i$-th neuron.  The original model \cite{gas:lom} was a first-of-its kind application of an HNN to identify bat echolocation calls in order to distinguish between a dataset of two cryptic, or morphologically similar, bat species \cite{bert:etal}. In this work, we shift from the relatively uncluttered soundscapes of the ultrasonic frequency range to a harder task, that of identifying specific calls of primates who vocalise in the more cluttered range audible to humans.

\subsubsection{Model 1.} 

Here, the two short representative call sound files are stored as shown in figure \ref{fig:flow1_activ}. It is worth noting that we use the terms store and train interchangeably as, while train is the more familiar term associated with CNN models, here and more strictly speaking, we are concerned with memory storage. The signals are shown as waveforms and spectrograms in figure \ref{fig:flow1_activ} (B). The fast Fourier transform (FFT) is then applied, which transforms these two signals from the time to the frequency domain.  We can see in figure \ref{fig:flow1_activ} (C) that alarms have a frequency composition dominating a higher frequency range than grumbles.  Alarms peak at $0.93$\,kHz while grumbles peak at $0.32$\,kHz. We tune the model to a frequency range containing the highest intensity/power peaks above a set threshold, adjust the number of neurons to ensure that memory capacity is not exceeded, and then fire/activate the HNN based on these peaks and frequency composition \cite{gas:lom}; see figure \ref{fig:flow1_activ} (D). Finally, in figure \ref{fig:flow1_activ} (E), the network activations representing each call signal are combined via Hebbian learning (equation \ref{eqn:Hebb}) to produce the trained model ready for activation by a new signal.  Positive and negative weights (neuron associations) are represented by green and red lines respectively whereas disassociated neurons (zero weights) are represented by transparent grey lines.  Once the network has been trained, the weights are fixed, and any pattern or, in our case, sound passed to the HNN layer will cause the network to evolve and converge to one of the two retrieval states (alarm or grumble) or to a spurious state \cite{gas:lom}, which is recorded as a model classification of UnID.  The labelled dataset of $1064$\,.WAV files (see section \ref{sec:data}) was then passed to the model, and over $63500$\,classifications recorded, one for each $1$\,second segment.

The raw sound files initially collected were $30$\,minutes long; for manual labelling purposes spectrograms were observed ($1024$-point FFT length and Hamming window function) over a $1$\,minute interval to identify whether there were calls within the signal.  For optimal detection, $1$\,second segments are passed to the HNN model for classification; presented with a $1$\,minute sound file, the network will record approximately $60$\,classifications which are then saved as a .csv file.  The model parameters include the full frequency range where peak frequencies are identified, the power threshold below which peaks are ignored, the segments of the signal which are passed to the HNN model and the number of neurons in the network.  These parameters can all be easily adjusted for the particular classification task.  For each experimental adjustment of parameters the network will be retrained and will generate a new set of classifications.  This takes a matter of seconds or minutes depending only on the amount of data to be processed, since training (i.e., sound storage) takes less than $10$ milliseconds. For model $1$ the frequency range used was from $0$ to $1.3$\,kHz, the normalised power threshold is $0.1$, the signal segment size is $1$\,second and the number of neurons in the network was set to $14$.

\subsubsection{Model 2.} 

The second model developed was trained on the two lemur calls (alarm and grumble) and also on a noise file representative of the sounds collected as the lemurs moved around and on the branch to which the recorder was attached.  This was to test the idea that we can use additional stored sounds to effectively filter out sounds that may affect accurate classification.  We should remember that the network will always converge to a retrieval state or a spurious state \cite{gas:lom} and that by storing a third signal, model 2 now has three retrieval states and spurious states.  In figure \ref{fig:flow3_activ} we represent model 2 training, and visualise the three stored signal representations of noise, grumble and alarm (B).  In (C) we observe the frequency composition of interest for each signal, with the noise signal peaking at approximately $0.07$\,kHz.  The number of neurons is increased to $34$ so as not to exceed the storage capacity of the network.  The new architecture and neuron activations, shown in figure \ref{fig:flow3_activ} (D), are combined via Hebbian learning to form the trained network represented in (E).  In model 2, once a signal is passed to the HNN layer, the network will converge to one of the retrieval or spurious states and the signal will be classified as either alarm, grumble, noise or UnID depending on which state the network converged to.  This information is then recorded to a .csv file containing the $1$\,second classifications.

\begin{table}[ht]
\caption{Classification reports for models $1$ and $2$.}
\begin{center}\footnotesize
 \begin{tabular}{l|cccc|cccc}
   \toprule
   Class & \multicolumn{4}{c|}{Model 1} & \multicolumn{4}{c}{Model 2} \\
   \cmidrule(lr){2-5}\cmidrule(lr){6-9}
    & Precision & \,Recall\, & \,F1\, & Support & Precision & \,Recall\, & \,F1\, & Support \\
   \midrule
   Grumble  & 0.53 & 0.81 & 0.64 & 203 & 0.80 & 0.80 & 0.80 & 203 \\
   Alarm & 0.92 & 0.69 & 0.79 & 32 & 0.90& 0.84& 0.87 & 32 \\
   Non-call & 0.90 & 0.97 & 0.93 & 1263 & 0.97 & 0.97 & 0.97 & 1263 \\
   \midrule\midrule
   & \multicolumn{3}{c|}{Overall Accuracy: 0.83} & 1498 & \multicolumn{3}{c|}{Overall Accuracy: 0.94} & 1498 \\
   \bottomrule
 \end{tabular}
\label{tab:CR}
\end{center}
\end{table}

\section{Results}\label{sec:results}
Here we present the performance metrics for each model (1 and 2) described in section \ref{sec:models} after the labelled dataset of $1064$ .WAV files is passed to each trained HNN.  Model 1 was trained on two black-and-white ruffed lemur calls (see section \ref{sec:data}) and therefore the HNN can converge to two retrieval states, classified as ``grumble'' and ``alarm'', or spurious states classified as ``UnID''.  Model 2 has been trained on a third signal which overlaps the frequency composition of the grumble signal. 
 Therefore, the HNN in this model can converge to three retrieval states, with the addition of a third class ``noise'', or spurious states.  Both models 1 and 2 are adapted from \cite{gas:lom}, tuned to the peak frequency range of the lemur vocalisations of interest, and set to provide classifications for each second of data within the raw sound files to be processed.  The models are tested on the labelled audio dataset (see section \ref{sec:data}) collected in the zoo enclosure housing three black-and-white ruffed lemurs.  The classification reports for both models are presented in table \ref{tab:CR}.  The metrics relate to the classification and identification of bouts of grumbles, alarms and non-calls.  As described in section \ref{sec:data}, grumble bouts were defined as being $2$\,seconds or more, alarm bouts $3$ seconds or more, and non-call bouts between $1$\,second and $60$\,seconds.

Our models yield classifications for each $1$\,second segment of data: a positive identification for a grumble bout therefore requires two consecutive detection records in the output .csv file (as demonstrated in table \ref{tab:Det}); for an alarm bout, three consecutive detection records are required; a non-call bout requires that between grumble bouts there is at least $1$\,second classified as ``alarm'', ``UnID'' or ``noise'', and between alarm bouts at least $5$\,seconds classified as ``grumble'', ``noise'' or ``UnID''.  In the use case described in this paper, where call monitoring can assess welfare or environmental interventions in captive settings, both recall and precision are relevant. It is important that a model can accurately identify the vocalisation in question (precision), and also a high proportion of instances of the vocalisation (recall). In table \ref{tab:CR} and very promisingly, we observe that model 1 gives high precision for alarm bout detection at $0.92$, and high recall for grumble bouts at $0.81$.  However, alarm bout recall of $0.69$ indicates that over $30\%$ of bouts are missed across the labelled dataset. The precision of grumble bout detection was low in model 1, with $47\%$ of the total grumble detections being false.  Model 2 shows consistent improvement across all metrics apart from alarm precision, and grumble recall; however, these remain high at $0.90$ and $0.80$ respectively.  As both precision and recall are relevant, and because the dataset is unbalanced due to the relative vocalisation rates of grumbles vs alarms, with the former being more common (see section \ref{sec:data}), an aggregated metric is of value to assess overall performance.  Here we use a weighted average or harmonic mean combining precision and recall, the F1-score, which is included in table \ref{tab:CR}.  We observe that the F1-scores for all bout types improve significantly once the noise signal is also included in the trained HNN.  The detection of non-call bouts is high across both models, but still shows significant improvement in precision in model 2.  One minute periods were chosen as the maximum length of a non-call bout to demonstrate just how precisely these models start to identify signal segments of interest.  When a different metric was used where non-call bouts could be any length across the dataset between alarm and/or grumble bouts (the longest in this dataset being over $2$\,hours), precision and recall for non-call bouts was still over $0.80$.  The overall accuracy for model 1 is $0.83$, and this increases to $0.94$ for model 2. Along with the improvements in recall of alarms and precision of grumbles, this provides strong evidence that model performance is improved by using both signals of interest and additional ``noise'' signals to improve detection accuracy.

\begin{figure}[ht]
\centering
\includegraphics[width=1.0\textwidth]{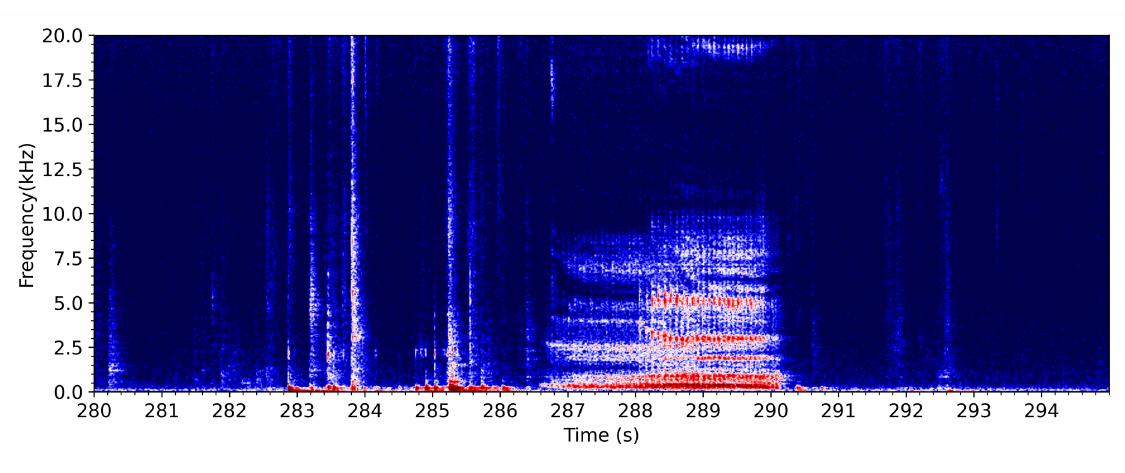}
\caption{Full spectrum visualisation of a grumble from $286.6$\,-\,$290.2$\,seconds as well as noises from lemur movement around the enclosure ($282$\,-\,$286$\,s). The spectrogram shows time (s), frequency (kHz) and amplitude as represented by colour intensity, with \textcolor{darkred}{\bf red} equating to high intensity/dB, and \textcolor{darkblue}{\bf blue} low intensity.
}\label{fig:det}
\end{figure}

To demonstrate this improvement we can use the example of a portion of one sound file containing a grumble and other noises made as the lemurs move around their enclosure. We would certainly recommend placement of the recorder away from apparatus accessible to the research subjects in future studies. However, given the data backlog \cite{ker:etal} it is not unusual to be presented with data collected for future analysis, and therefore it is important to develop methods to deal with unwanted sounds without losing any signal information, including those unexpected sounds with similar frequency composition to the calls of interest. The spectrogram in figure \ref{fig:det} represents a section of .WAV file containing sounds due to movement and then a grumble bout between $286.6$\,-\,$290.2$\,seconds. Dark red represents the frequencies with highest power and dark blue the lowest. Table \ref{tab:Det} shows the classifications associated with this section of the .WAV file.  We observe that while both models successfully identify the grumble bout (recall that two or more consecutive grumble detections are required to register a grumble bout), model 1 incorrectly detects a grumble bout between $282$\,-\,$284$\,seconds.  This portion of the signal actually contains noise due to movement and has overlapping frequency content with the stored grumble within the HNN. Model 2 classifications, as shown in table \ref{tab:Det}, demonstrate that once the noise file is also stored in the HNN, the detection of grumble bouts becomes more accurate in terms of duration, and that there is no longer a false detection due to the recorded sounds of movements.

\begin{table}[ht]
\caption{Model comparison for classifications over $1$\,second segments from the raw sound file in figure \ref{fig:det}, showing classification start time\,(s) and classification labels. U: UnID; G: Grumble; N: Noise; \textcolor{redfill}{\bf Red fill}: false positive; \textcolor{mygreen}{\bf Green fill}: true positive.}
\begin{center}\footnotesize
 \begin{tabular}{ p{2cm}||p{0.55cm}|p{0.55cm}|p{0.55cm}|p{0.55cm}|p{0.55cm}|p{0.55cm}|p{0.55cm}|p{0.55cm}|p{0.55cm}|p{0.55cm}|p{0.55cm}|p{0.55cm}|p{0.55cm}|p{0.55cm}|p{0.55cm}}
    \toprule  Classifcation time (s)&280&281&282&283&284&285&286&287&288&289&290&291&292&293&294\\
    \midrule\midrule
    Model 1 &U&U&\cellcolor{redfill}G&\cellcolor{redfill}G&\cellcolor{redfill}G&U&U&\cellcolor{greenfill}G&\cellcolor{greenfill}G&\cellcolor{greenfill}G&U&U&U&U&U\\
    \midrule
    Model 2 &N&N&U&U&N&U&\cellcolor{greenfill}G&\cellcolor{greenfill}G&\cellcolor{greenfill}G&\cellcolor{greenfill}G&U&N&N&N&N\\
    \bottomrule
 \end{tabular}
\label{tab:Det}
\end{center}
\end{table}

\section{Discussion}\label{sec:discussion}
Here we discuss how further improvements to the results outlined in section \ref{sec:results} could be obtained, and how the lightweight model could be repurposed for other tasks and questions. Although the aim of this work is to demonstrate that AI models of this type \cite{gas:lom} are successful for bioacoustic event detection in captive settings, we are optimistic that further refinements will continue to improve model performance.  For this particular classification problem, further refinements could include storing additional noise signals in the HNN.  The majority of alarm false positives occurred during cleaning (the scrape of a shovel along the floor for example) and a particular crunch while lemurs were eating.  Alarm false negatives, which affect the recall metric in table \ref{tab:CR}, often occurred due to the AM's microphone being overloaded by alarm calls made in close proximity to it. Grumble bout detections were similarly impacted by recorder placement, as not only did the AM collect signals due to movement along the branch to which it was attached, but also, being within easy reach, the sound of direct contact with the AM.  The stored noise in model 2 reduced grumble false positives and improved precision from $0.53$ to $0.80$ (table \ref{tab:CR}).  However, when noise due to recorder placement overlapped a grumble bout, the frequency composition during a $1$\,second segment changed sufficiently to mean that while a single detection may be captured during the bout, two consecutive detections were not in model 2. Both types of call detection were further impacted by the method chosen to label the data; although the method of splitting the audio data into $1$\,minute .WAV files optimised the manual labelling process, some detections were lost due to a call bout straddling a split i.e, call bouts starting and ending at the end of one .WAV file and the beginning of another respectively.

The associative memory models demonstrated here are extremely efficient, taking less than $10$ milliseconds to train, and are able to rapidly classify the one second segments of data passed to the HNN.  A MacBook Air, with $16$\,GB RAM and an M3 chip, ran model 2 in the background while other applications were in use; over $5.5$\,hours of audio data was processed and classified per minute, equating to around $340$\,detections and classifications per second.  We have discussed improvements to the model by storing further noise samples to the HNN, although prior to that we would recommend examination of the spurious states to better understand misclassification.  The resolution of the model can also be tuned by adjusting the 'length of segment' parameter to generate more (or less) detailed classifications across the datasets used, depending on user requirements. The position and type of the recording device should be planned carefully before deployment to avoid unnecessary ``polluting'' sounds being recorded; however, we are aware that prior to audio data collection researchers may not be aware of which sounds are significant.  While here we are concerned with specific primate vocalisations, a count of the number of door slams, passes across a particular piece of apparatus, urination events, and how soon after placing food in an enclosure before eating commences could all be of relevance to a welfare intervention study and could all be stored to a new HNN model.  It is also recommended that, rather than daily setup and removal of recording equipment, a continuous recording approach is adopted, ideally with equipment that can be started and paused remotely.  Acoustic data is often an under-utilised resource in conservation settings that historically have been dependant on visual survey methods.  Although PAM is increasingly used, sustainable solutions to automating the processing of the large datasets collected are urgently required \cite{ker:etal}. The promising results documented here demonstrate that these lightweight associative memory models based upon a Hopfield neural network have broad applicability across a wide frequency range and can be quickly repurposed. They have no heavy-duty hardware requirements, are not restricted by the length of the audio files collected, can process large quantities of data rapidly and can be adapted swiftly in the field to different conservation questions and scenarios.  Here, a model developed to classify bat echolocation calls in an uncluttered high frequency soundscape \cite{gas:lom}, was quickly repurposed to track primate vocalisations collected in a noisy captive setting.

\section{Conclusions and Future Work}\label{sec:conc}

In this work, we demonstrate an adaptable and robust proof-of-concept for the automated monitoring of captive primate vocalisations as a conservation and welfare aid.  The AI model used is an alternative to the CNNs currently dominating the research field \cite{ker:etal,ras:sto:bri,cau:fav:mar:rey,bat:etal}. This lightweight associative memory model incorporating Hopfield neural network architecture, originally developed to detect two cryptic bat species \cite{gas:lom}, has been adapted to the cluttered lower frequency range of primate vocalisations.  No costly conversion to images or large prelabelled datasets are required for training (unlike the popular CNN models), just single sound files of $1$\,-\,$2$\,seconds in length.  We create a labelled dataset of endangered black-and-white ruffed lemur calls for model evaluation by identifying and labelling bouts of alarms, grumbles and non-calls of interest, collected on an AudioMoth in 2023 in a zoo enclosure home to three lemurs. Examples of the vocalisations of interest, grumble and alarm bouts, were selected from the first day of recordings and used to train the model, alongside an additional representative noise file.  Model 2, which was trained on all three stored sounds, performed well with overall accuracy of $0.94$, an increase in accuracy from model 1, indicating that the storage of the additional noise file successfully reduced false positives and false negatives without any loss of signal information.  While the files used for storage/training were from a one hour period on day one of recording, the model was able to generalise and perform well on the whole dataset which was collected across one month. The model produced is lightweight enough to train in less than $10$\,milliseconds including all preprocessing of the training data and perform around $340$\,classifications per second, meaning that over $5.5$\,hours of recorded audio files can be processed and classified per minute. In contrast, manual labelling of $5.5$\,hours of data took over $100$\,times longer at approximately $1.7$\,hours.

We suggest a number of factors for consideration before the selection and placement of audio recorders to help increase accuracy, but also acknowledge that sounds other than vocalisations could be of value to acoustic studies in captive settings. These sounds could easily be stored to create a new trained model. Due to the speed of training, completely new or adapted models can be built and then used to classify an entire dataset, the classification process taking only a few minutes; identifying the new sounds of interest being the more time-costly element to the redeployment. The models developed here aim to complement conservation goals around sustainability by providing low resource intensity solutions to PAM dataset processing backlogs. They can provide uniquely adaptable solutions to new problems where large labelled datasets are not available.  Furthermore, we aim to create transparent, explainable AI solutions where every event detection whether true or false, is understandable for both the AI expert and the field worker, thus improving uptake and further impacting vital conservation efforts. While these lightweight models could certainly be deployed on real-time detection devices, we do not aim to completely replace the time a field-worker spends with their data nor their experience, rather to reduce data-to-insight turnaround times and accelerate decision making.  By adopting a human-in-the-loop approach and working alongside conservationists, ecologists and zoo staff we aim to use these novel models to enhance their skills, detect new patterns and gather new insights about a complex acoustic world often overlooked by visually oriented humans.

\section{Acknowledgements}\label{sec:ack}
This project is generously supported by an OpenBright Award, an Invest to Grow studentship and RIF4 Biosciences Research Project funding (bioacoustic equipment, software analysis and LW salary) from the University of Wolverhampton.  The authors would also like to thank staff at Dudley Zoo and Castle (UK) for their expert insight and assistance, and Jessica Gwilliams for data collection.

%
%


\begin{thebibliography}{6}
%

\bibitem{bak}
Bakker, K.J.: The Sounds of Life: How Digital Technology is Bringing Us Closer to the Worlds of Animals and Plants.
Princeton University Press, Princeton, NY (2022)

\bibitem {ker:etal}
Kershenbaum, A., Akçay, Ç., Babu-Saheer, L., Barnhill, A., Best, P., Cauzinille, J., Clink, D., Dassow, A., Dufourq, E., Growcott, J., Markham, A., Marti-Domken, B., Marxer, R., Muir, J., Reynolds, S., Root-Gutteridge, H., Sadhukhan, S., Schindler, L., Smith, B.R., Stowell, D., Wascher, C.A.F., Dunn, J.C.: Automatic detection for bioacoustic research: a practical guide from and for biologists and computer scientists. 
In: Biol Rev, vol. 100(2), pp 620-646, (2024). 
\url{doi.org/10.1111/brv.13155}

\bibitem{mar:fab:aub}
Marchal, J., Fabianek, F., Aubry, Y.: Software performance for the automated identification of bird vocalisations: the case of two closely related species.
In: Bioacoustics, vol. 31(4), pp 397--413, (2022).
\url{doi.org/10.1080/09524622.2021.1945952}

\bibitem{tab:etal}
Tabak, M.A., Murray, K.L., Reed, A.M., Lombardi, J.A., Bay, K.J.: Automated classification of bat echolocation call recordings with artificial intelligence.
In: Ecological Informatics, vol. 68, pp 101526, (2022).
\url{doi.org/10.1016/j.ecoinf.2021.101526}

\bibitem{ras:sto:bri}
Rasmussen, J.H., Stowell, D., Briefer, E.F.: Sound evidence for biodiversity monitoring.
In: Science, vol. 385(6705), pp 138--140, (2024).
\url{doi.org/10.1126/science.adh2716}

\bibitem{cau:fav:mar:rey}
Cauzinille, J., Favre, B., Marxer, R., Rey, A.: Applying machine learning to primate bioacoustics.
In: Review and perspectives. American Journal of Primatology, vol. 86(10), (2024). 
\url{doi.org/10.1002/ajp.23666}

\bibitem{bat:etal}
Batist, C. H., Dufourq, E., Jeantet, L., Razafindraibe, M. N., Randriamanantena, F., Baden, A. L.: An integrated passive acoustic monitoring and deep learning pipeline for black-and-white ruffed lemurs (Varecia variegata) in Ranomafana National Park, Madagascar. 
In: American Journal of Primatology, vol. 86 (2024). \url{doi.org/10.1002/ajp.23599}

\bibitem{cla:dun}
Clark, F.E, Dunn, J.C.: From Soundwave to Soundscape: A Guide to Acoustic Research in Captive Animal Environments. 
In: Front Vet Sci., vol. 9, (2022). 
\url{doi.org/10.3389/fvets.2022.889117}


\bibitem{iuc}
IUCN.: The IUCN Red List of Threatened Species. Version 2024-2 (2024) 
\url{https://www.iucnredlist.org}


\bibitem{hil:etal}
Hill, A.P., Prince, P., Snaddon, J.L., Doncaster, C.P., Rogers, A.: AudioMoth: A low-cost acoustic device for monitoring biodiversity and the environment.
In: HardwareX, vol. 6 (2019).
\url{doi.org/10.1016/j.ohx.2019.e00073}


\bibitem{gas:lom}
Gascoyne, A., Lomas, W.: First-of-its-kind AI model for bioacoustic detection using a lightweight associative memory Hopfield neural network.
In: Ecological Informatics, vol. 91, pp 103382, (2025).
\url{doi.org/10.1016/j.ecoinf.2025.103382}

\bibitem{hop}
Hopfield, J.J.: Neural networks and physical systems with emergent collective computational abilities.
In: Proceedings of the National Academy of Sciences, USA, vol. 79, pp. 2554--2558 (1982)
\url{doi.org/10.1073/pnas.79.8.2554}


\bibitem{isi}
Ising, E.: beitrag zur theorie des ferromagnetismus. 
In: Z. Phys. vol. 31(1), pp. 253–258 (1925)


\bibitem{hop:tan}
Hopfield, J. J., Tank, D. W.: "Neural" computation of decisions in optimization problems.
In: Biological cybernetics, vol. 52, pp. 141--152 (1985).
\url{doi.org/10.1007/BF00339943}


\bibitem{rod:etal}
Rodden, E., Gascoyne, A., Naughton, L., Brennan, J., Parkes, A.: Transient Chaotic Neural Network with Negative Self-feedback Memory for Continuous Optimisation Problems. 
In: Arai, K. (eds) Proceedings of the Future Technologies Conference (FTC), vol. 1. Lecture Notes in Networks and Systems, vol. 1154. Springer, Cham. (2024).
\url{doi.org/10.1007/978-3-031-73110-5_19}

\bibitem{che:aih}
Chen, L., Aihara, K.: Chaotic simulated annealing by a neural network model with transient chaos.
In: Neural Networks, vol. 8(6), pp. 915–930 (1995). 
\url{doi.org/10.1016/0893-6080(95)00033-V}
 
\bibitem{xu:zha}
Xu, Y., Zhao, T.: Chaotic neural network with nonlinear function self-feedback.
In: Proceedings of the 33rd Chinese Control Conference, Nanjing, China, pp. 5075-5079 (2014).
 

\bibitem {elw:vag}
Elwell, E. J., Vaglio, S.: The Scent Enriched Primate. 
In: Animals, vol. 13(10), pp 1617, (2023). 
\url{doi.org/10.3390/ani13101617}

\bibitem{fon:etal}
Fontani, S., Glendewar, G., Cowen, R., Callagan, G., Costantini, A.B., Elwell, E., Dubreuil, C., Palframan, M., Vaglio, S.: Novel Scent Enrichment Enhances Socio-Sexual and Olfactory Behaviors in Zoo-Housed Gentle Lemurs. 
In: American Journal of Primatology, vol. 87(1) (2025).
\url{doi.org/10.1002/ajp.23716}


\bibitem{goo}
Goodwin, W.: VIDEO + AUDIO: What do lemurs sound like?
Duke Lemur Centre (2019).
\url{https://lemur.duke.edu/video-audio-what-do-lemurs-sound-like/}


\bibitem{bat:raz:ran:bad}
Batist, C.H., Razafindraibe, M.N., Randriamanantena, F., Baden, A.L.: Bioacoustic characterization of the black-and-white ruffed lemur (Varecia variegata) vocal repertoire.
In:Primates, vol. 64(6), pp. 621–635 (2023). \url{doi.org/10.1007/s10329-023-01083-8}

\bibitem{heb}
Hebb, D.O.: The Organization of Behavior: A Neuropsychological Theory, 1st ed.
Wiley, Oxford (1949)

\bibitem {bert:etal}
Bertran, M. and Alsina-Pagès, R. M., Tena, E.:Pipistrellus pipistrellus and Pipistrellus pygmaeus in the Iberian Peninsula: An Annotated Segmented Dataset and a Proof of Concept of a Classifier in a Real Environment. 
In: Applied Sciences, vol. 9, (2019). 
\url{doi.org/10.3390/app9173467}



\end{thebibliography}
\end{document}